# Deep learning-based shot-domain seismic deblending

Jing Sun[1,2], Song Hou[3], Vetle Vinje[2], Gordon Poole[3], Leiv-J Gelius[1]

[1]University of Oslo, Department of Geosciences, Sem Sælandsvei 1, Oslo NO 0316, Norway.

[2]CGG Services Norway AS, Lilleakerveien 6A, Box 43, Lilleaker NO 0283, Norway.

[3]CGG Services, R&D Crompton Way, Manor Royal Estate Crawley, West Sussex RH10 9QN, UK.

## ABSTRACT

To streamline fast-track processing of large data volumes, we have developed a deep learning approach to deblend seismic data in the shot domain based on a practical strategy for generating high-quality training data along with a list of data conditioning techniques to improve performance of the data-driven model. We make use of unblended shot gathers acquired at the end of each sail line, to which the access requires no additional time or labor costs beyond the blended acquisition. By manually blending these data we obtain training data with good control of the ground truth and fully adapted to the given survey. Furthermore, we train a deep neural network using multi-channel inputs that include adjacent blended shot gathers as additional channels. The prediction of the blending noise is added in as a related and auxiliary task with the main task of the network being the prediction of the primary-source events. Blending noise in the ground truth is scaled down during the training and validation process due to its excessively strong amplitudes. As part of the process, the to-be-deblended shot gathers are aligned by the blending noise. Implementation on field blended-by-acquisition data demonstrates that introducing the suggested data conditioning steps can considerably reduce the leakage of primary-source events in the deep part of the blended section. The complete proposed approach performs almost as well as a conventional algorithm in the shallow section and shows great advantage in efficiency. It performs slightly worse for larger traveltimes, but still removes the blending noise efficiently.

## INTRODUCTION

Within the field of seismic exploration, survey designs have been made denser and larger in an attempt to improve imaging quality. The increased amount of recorded seismic data in combination with the pressure to deliver processing results within a limited time frame represent a major challenge to the industry. Thus, there is a demand for the development of more time-efficient processing approaches. Deep learning (DL), a subfield of machine learning in artificial intelligence, allows computational networks that are composed of multiple processing layers to learn representations of data with multiple levels of abstraction (LeCun et al., 2015). The essence of DL is to learn and make predictions from data and extract important correlations about physical processes characterizing large data sets. Ideally, machine learning represents the science that enables computers to learn without explicit programming. The main idea being that general algorithms exist that can be used to identify patterns in a wide range of data sets without the need to write specific code tailored for each problem.

Within the field of seismic processing, application of DL has drawn much attention. Many attempts have been made to solve the trace interpolation problem (Siahkoohi et al., 2018; Oliveira et al., 2018; Mandelli et al., 2018; Wang et al., 2018; Kaur et al., 2019; Wang et al., 2020; Larsen Greiner et al., 2020). The study of DL within seismic noise attenuation has also started to develop. Li et al. (2018) and Kaur et al. (2020) investigated the attenuation of ground-roll noise. Yu et al. (2019) investigated the potential application of a convolutional neural network (CNN) to attenuate random noise, linear noise and multiples. Slang et al. (2019) and Sun et al. (2020) demonstrated the use of a U-Net to remove seismic interference noise from marine field data. Besides applications within noise attenuation and interpolation, DL has also already proved to be useful in various other areas such as velocity model building (Araya-Polo et al., 2018; Richardson, 2018) and geologic feature identification of faults (Araya-Polo et al., 2017; Huang et al., 2017; Guo et al., 2018; Wu et al., 2019; Xiong et al, 2018; Zhao and Mukhopadhyay, 2018), salt bodies (Waldeland et al., 2018; Shi et al., 2018; Di et al., 2019) and channels (Pham et al., 2019).

Although not yet fully explored, the use of DL within the field of seismic deblending is also underway. Most studies have focused on deep neural network (DNNs) with supervised learning. A major challenge for such methods is the lack of ground truth in the training data (e.g. perfectly deblended data) in case of blended-by-acquisition data. Training on synthetic data and application on field data is not optimal due to the high complexity of the latter. Alternative use of processed results output from a conventional workflow as the ground truth, is both expensive and not ideal in terms of data quality. In case of larger jitter times, the blending noise (data from non-first sources) is much stronger and masks the primary energy associated with the first source. Thus, leading to a processing challenge due to the poor signal-to-noise ratio (SNR). Besides, earlier attempts of using DL to deblend seismic data have used data sorted in common-offset gathers (COG) or common-receiver gathers (CRG). Use of such a sorting strategy breaks the coherency of the contributions from the non-first sources and transforms it to the alternative problem of attenuating incoherent noise.

Slang et al. (2019) and Sun et al. (2020) investigated the use of a CNN to deblend seismic field data in the common channel domain. In these studies, deblended data obtained from the use of a conventional algorithm showing only weak blending noise were used as the ground truth. A time-square ($t^2$) gain function was applied before these data were manually blended. The trained network gave promising results, but it is important to realize that the use of $t^2$ amplitude-scaling leads to a higher SNR than normally encountered in a real field case and makes deblending easier. Richardson and Feller (2019) proposed the use of adjacent COGs as input to a U-Net approach to deblend marine field data. Baardman and Hegge (2020) proposed the initial use of a CNN to identify the blended and unblended parts of the data. A second CNN having many similarities with that of a classification network was then employed to deblend seismic data in the common receiver domain. The authors considered both synthetic and field data in their study. Zu et al. (2020) proposed an iterative CNN-based workflow for deblending in case of a two-source simultaneous shooting. Two different field data tests, employing data sorting in respectively the common receiver domain and the common offset domain were presented.

In contrast to previous investigations, we propose here a DL approach to deblend seismic data in the acquisition domain. Since both the quality and availability of training data are important issues for a DNN with supervised learning, a new strategy for generating training data is first proposed. In addition, we present a collection of suggested data preparation steps that improves the deblending accuracy but barely change the cost of application. Each of these data conditioning steps represents minor improvements but when applied collectively leads to a refined result.

This paper is organized as follows. First, an introduction to blended acquisition and a description of the specific field survey used for demonstration are given. After that, we discuss existing methods to generate training data for DNN-based deblending and propose a novel practical solution. Then, a sample of suggested data conditioning steps to improve the deblending accuracy of the network is presented. In the section to follow, we apply the proposed approach to field-blended data acquired from a source-over-streamer blended acquisition and compare the deblended results after stacking with a conventional deblending algorithm. Finally, a discussion and set of conclusions are given.

## BLENDED ACQUISITION AND TOPSEIS

In a typical blended acquisition, sources are fired with overlapping time differences including a small random jitter. For convenience, we define a series of notations to represent data from a blended acquisition with arbitrary number of sources and streamers. Note that a shot gather represents data associated with one source and one streamer. Let the shot gather corresponding to the $k$th shooting of source number $i$ and received by the $j$th streamer be represented as $\boldsymbol{S}_{i,k}^{j}$ (cf. Figure 1). Further, if we consider a fixed source $i$ and a streamer $j$, two adjacently fired shot gathers to $\boldsymbol{S}_{i,k}^{j}$ (one on each side) can be represented as $\boldsymbol{S}_{i,k-1}^{j}$ and $\boldsymbol{S}_{i,k+1}^{j}$ respectively. Data corresponding to the last shot point of a sail line is called a last shot gather. For a blended survey with $J$ streamers, $J$ last shot gathers can be obtained for each sail line and they are all unblended from subsequent shot activations.

Field data used for demonstration purposes in this investigation are taken from a source-over-streamer blended acquisition conducted across the Castberg field in the Barents Sea (Vinje and Elboth, 2019; Poole et al., 2020). The special acquisition technique is named TopSeis due to the location of the source on top of the seismic spread. The actual acquisition geometry is shown in Figure 1, where five source arrays (referred to as Top-Sources) with a nominal 3.0 s shooting rate and a shot-to-shot dithering of maximum ± 200ms were located approximately 30 m above the deep streamers. The shooting rate of 3.0 s leads to a dense shot sampling which means that seismic energy from different shots will be blended, termed “blended-by-acquisition”. The source towed by the streamer vessel (referred to as the Front-Source) was activated at every 6th Top-Source, with a ± 400ms dithering. Thus, adding additional Front-Source energy to every 6th Top-Source record. Note that prior to the deblending the direct source-to-receiver waves from the Top-Sources have been removed by a separate set of processing steps. In addition the Front-Source energy was removed from the blended-by-acquisition data set before being used in this study.

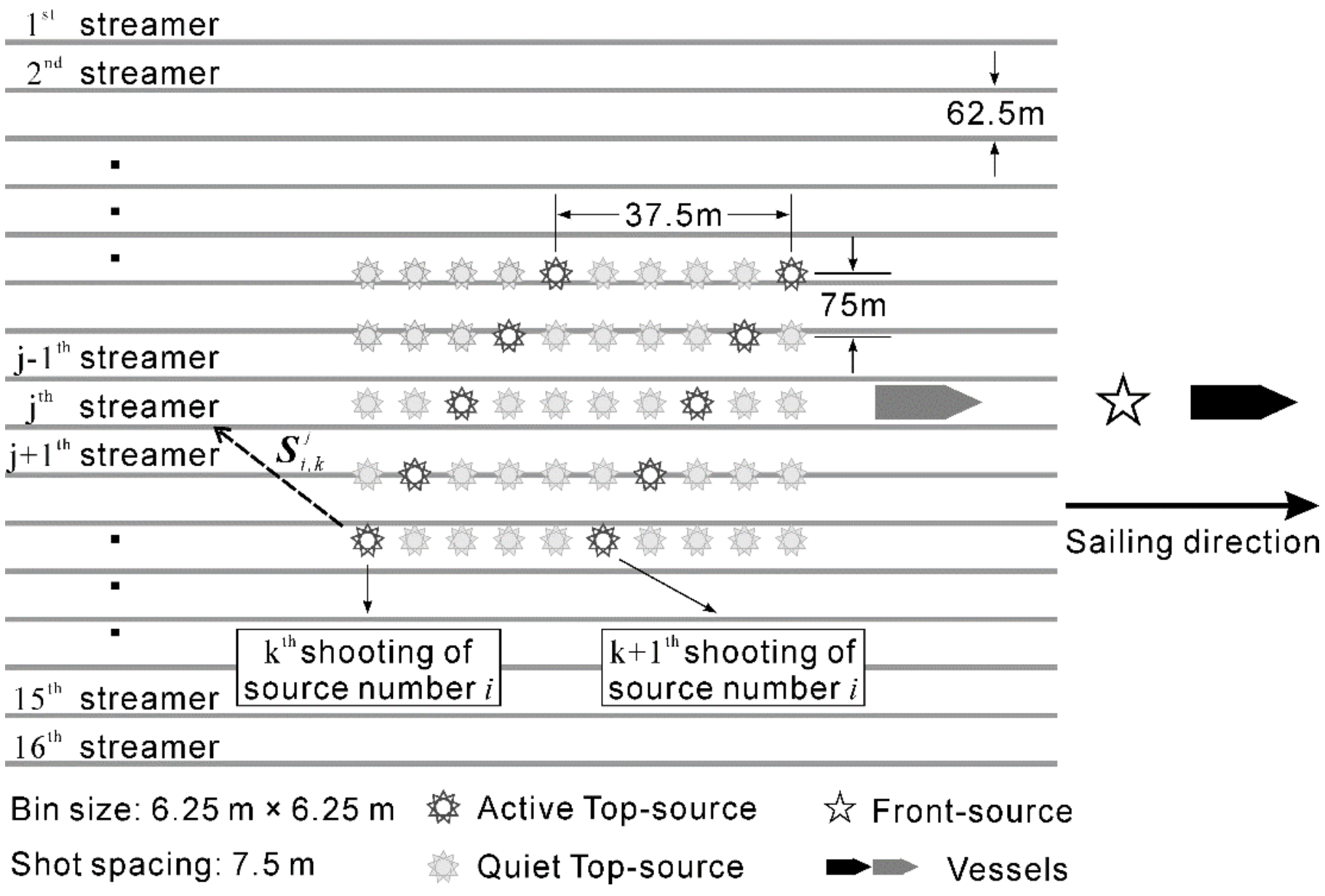

Figure 1: Lay-out of the source-over-streamer blended acquisition survey conducted in the Barents Sea.

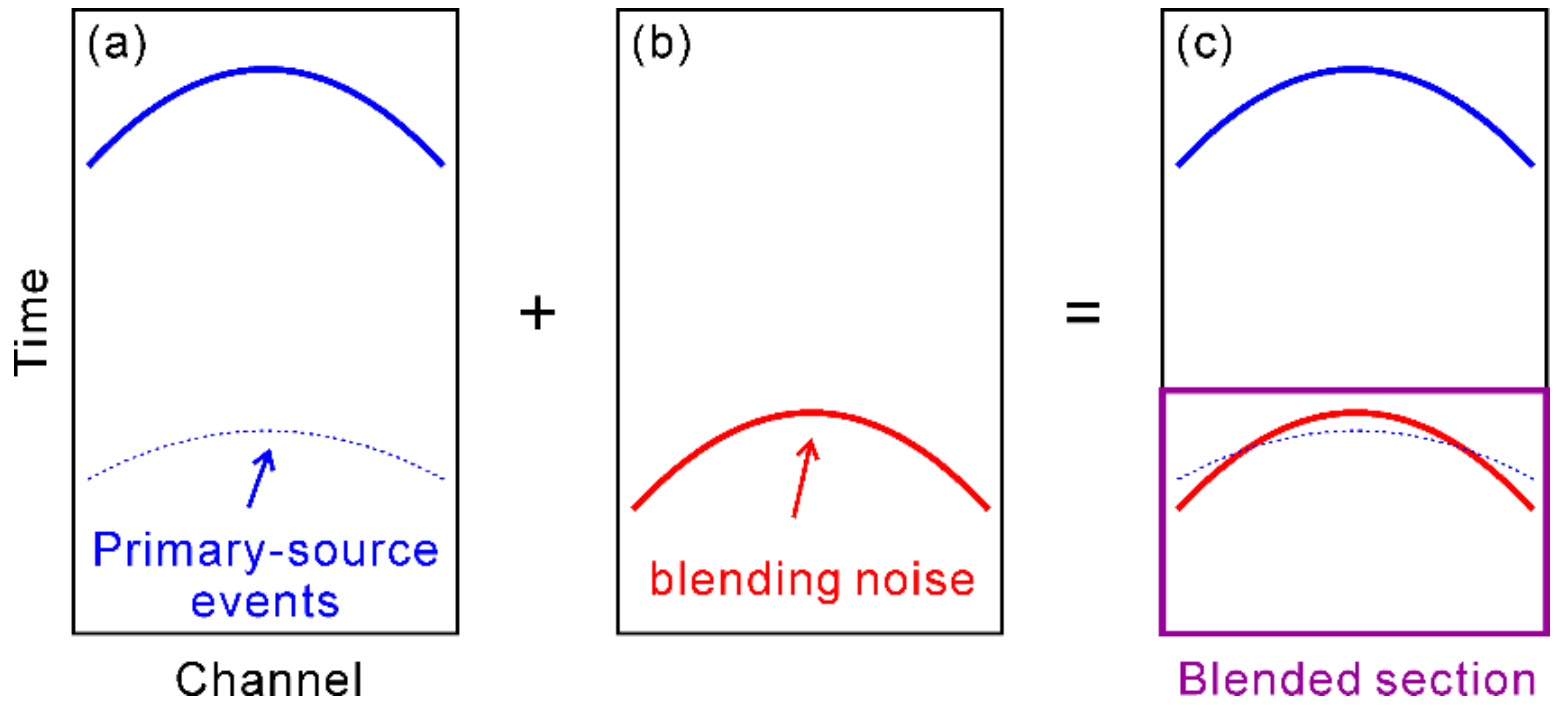

Figure 2: Schematics of a blended source-over-streamer shot gather. (a) Primary-source shot gather, (b) shot gather from the overlapping source and (c) blended shot gather. The purple box shows components of the blended section.

An example of a blended source-over-streamer shot gather is schematically shown in Figure 2. Figure 2a represents the primary-source gather and with Figure 2b illustrating the blending noise that is the corresponding shot gather from the overlapping source. The purple box in Figure 2c represents the blended section. The blue dotted curve in the blended section represents the primary-source events to be restored. To save computational memory and reduce the training time, only the blended sections were fed into the network. Figures 3a and 3b show respectively a typical blended-by-acquisition shot gather and an unblended last shot gather from the source-over-streamer blended acquisition. Figure 4 shows a typical example of the SNR in the blended section. The amplitude of the blending noise is more than one order of magnitude higher than that of the primary-source events.

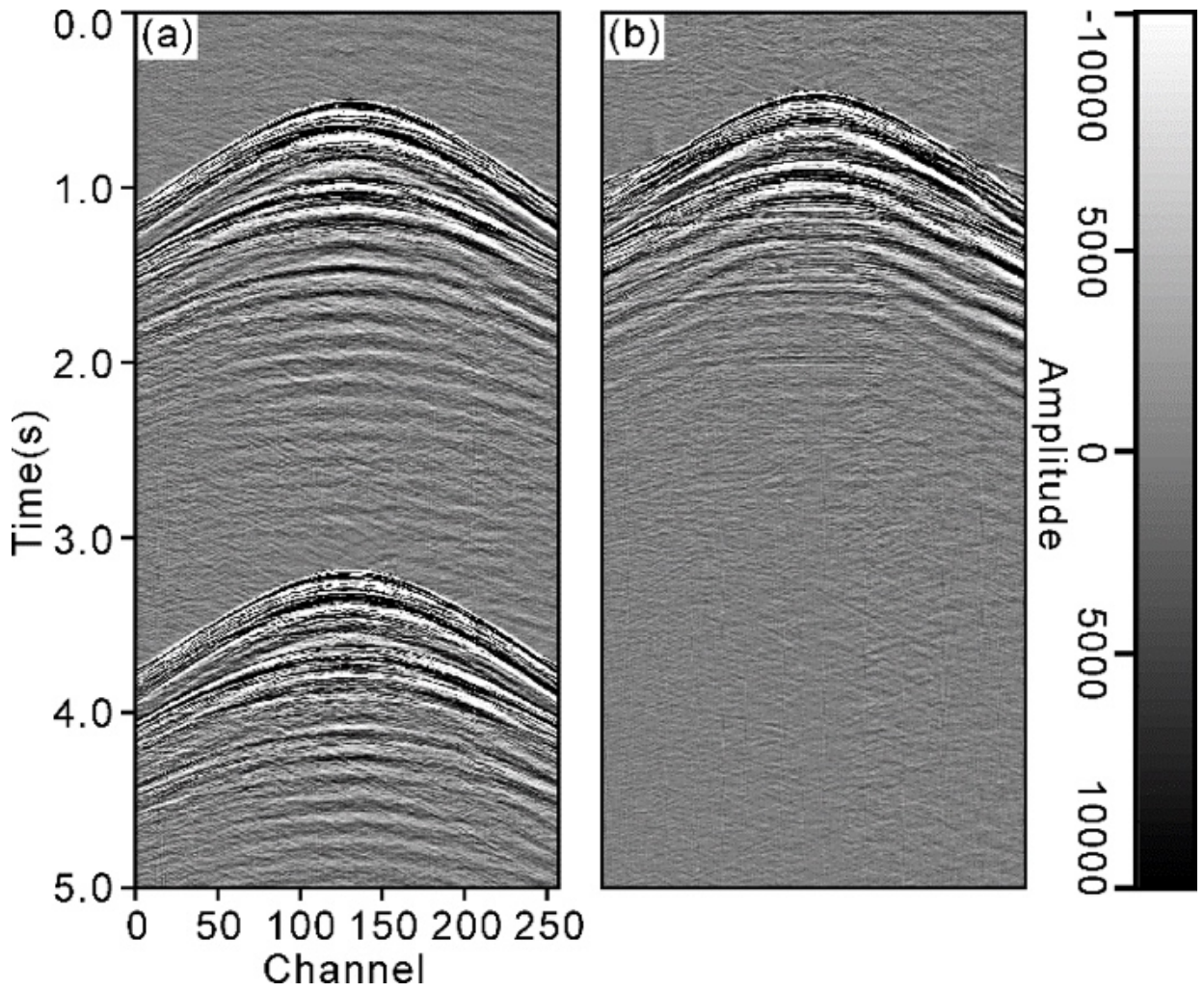

Figure 3: Example of (a) blended-by-acquisition source-over-streamer shot gather and (b) typical last shot gather from a source-over-streamer blended acquisition.

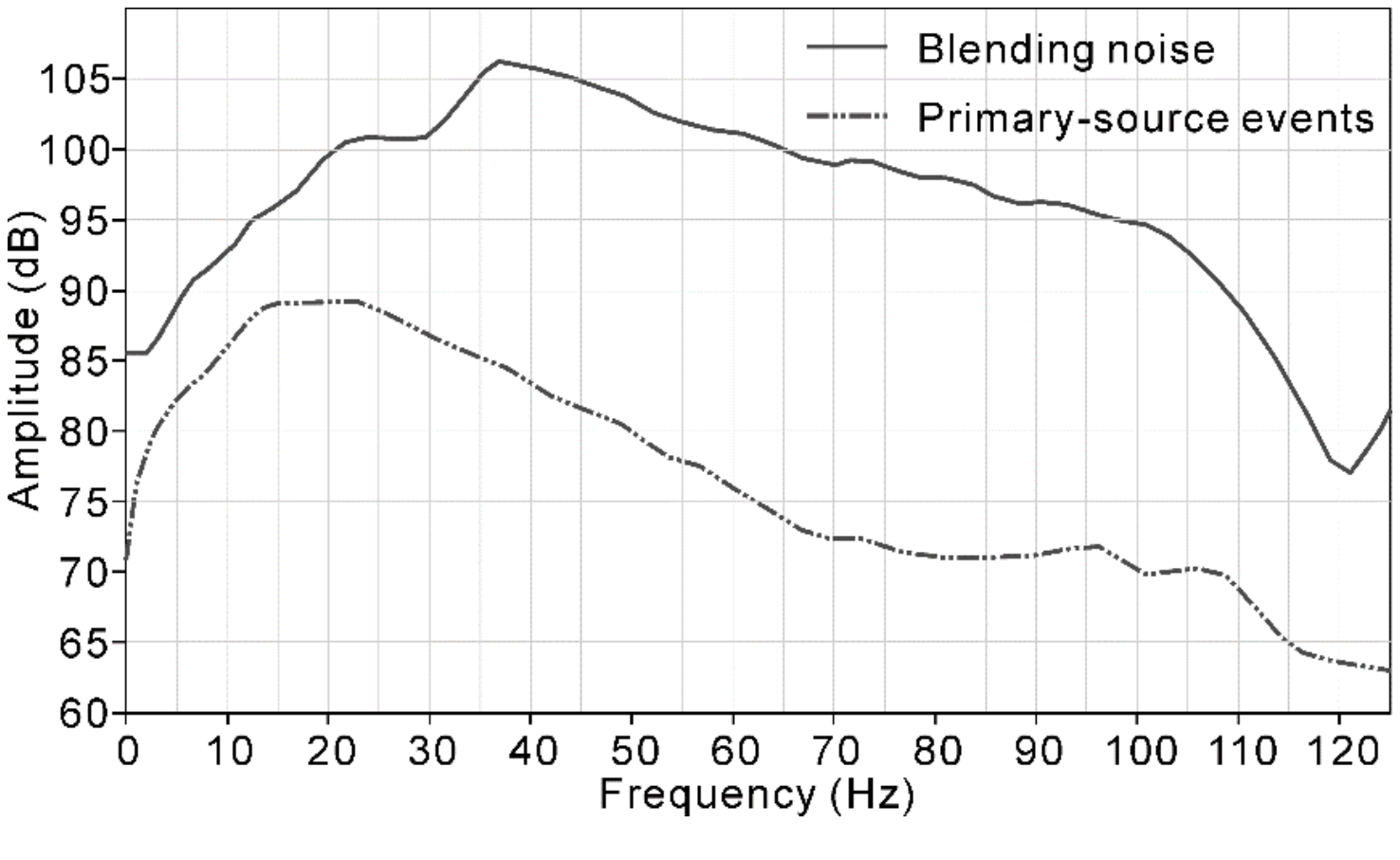

Figure 4: Example of SNR in the blended section of a typical blended-by-acquisition source-over-streamer shot gather from the Barents Sea.

## ARCHITECTURE OF THE EMPLOYED DNN

In this study, we applied a DNN architecture based on U-Net (Ronneberger et al., 2015). We tested different hyper-parameters and found they were not very sensitive to the DNN model accuracy compared to the training data sets. U-Net was initially proposed to solve semantic segmentation tasks, representing an encoder-decoder-structure with skip connections built in a U-shape inspired by the fully convolutional neural network of Long et al. (2015). As shown in Figure 5, the building block of the encoder consists of convolutional layers employing a typical filter size of 3×3 and a ReLU activation function (dark blue arrows). In addition, introduction of max pooling layers yield a multilevel, multi-resolution feature representation. The max pooling operation is represented by red arrows where the stride is 2 and the pool size is 2×2 (i.e. data are downscaled to half size in both spatial dimensions). The corresponding building block of the decoder employs transposed convolutional layers with a stride 2 represented by green arrows to up-sample low-resolution features describing large scale structures to full resolution feature maps. The skip connections between the encoding path and the decoding path employing a concatenation operation (grey arrows) ensure information fusion between high- and low-level information and can thereby improve the accuracy of the network performance. No activation function is applied at the last layer of the network.

We train the DNN by minimizing the mean squared error (MSE) between the ground truth and its prediction over all patches in the training data set which can be expressed as,

$$L(W, B) = \text{argmin} \sum \left(T - \hat{T}\right)^2, \tag{1}$$

where $T$ represents the prediction of the model, $\hat{T}$ represents the ground truth, $W$ represents the weights and $B$ represents the bias. To solve this optimization problem, we employed the Adam optimization algorithm (Kingma and Ba, 2014) on batches of 4 patches.

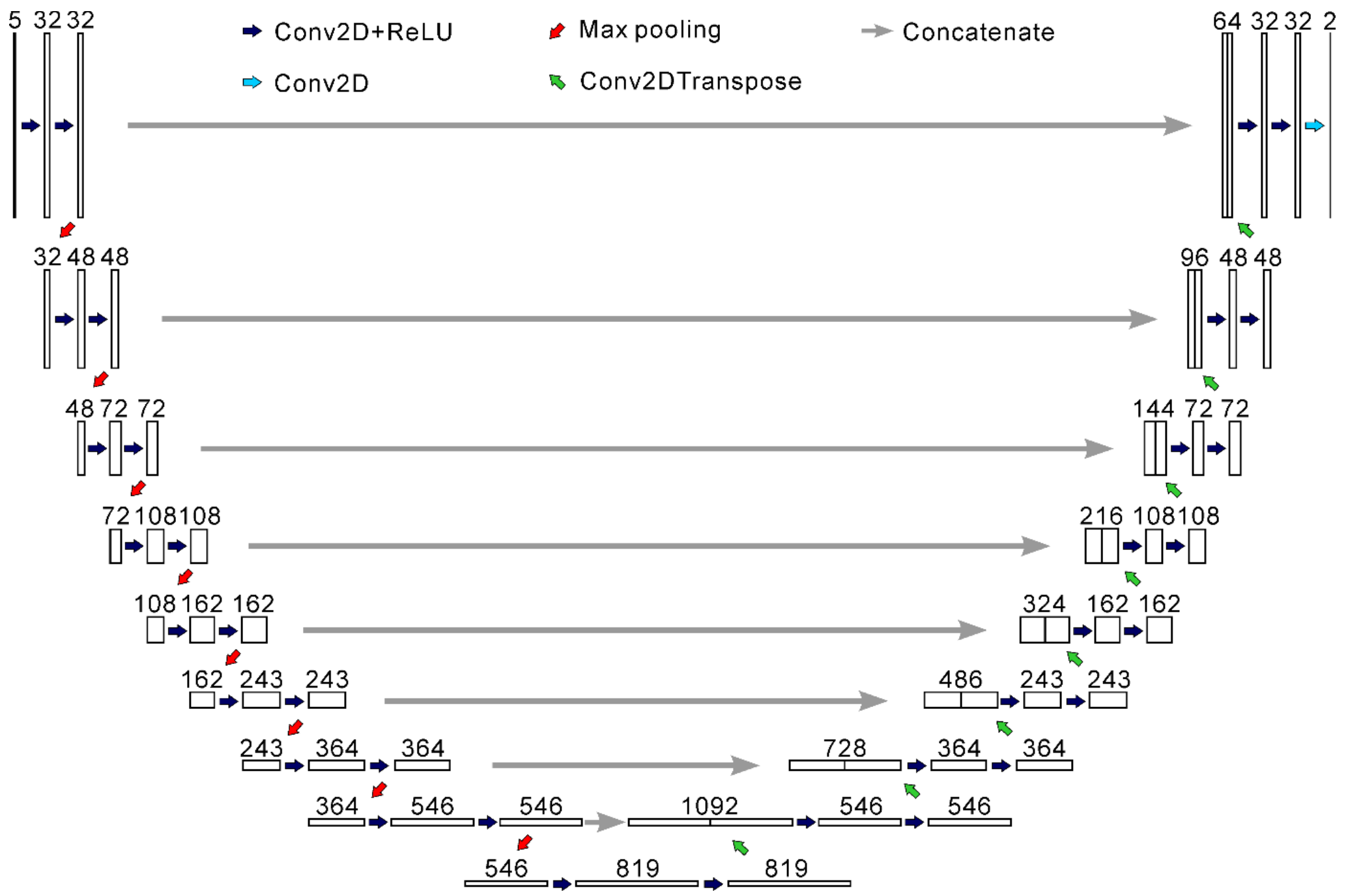

Figure 5: Architecture of the employed DNN. Each rectangle represents a collection of feature maps from the previous operation, and the numbers above (e.g. 32, 48, 72 and 108) represent the number of feature maps.

## A NEW STRATEGY TO GENERATE TRAINING DATA WITH HIGH QUALITY AND AVAILABILITY

DNN with supervised learning aims to establish the relation between input example data and the desired output or ground truth. The quality and availability of training data are therefore critical issues regarding the effectiveness of the training process and the accuracy of the trained predictions. Firstly, the quality of the ground truth data employed in the training process should be as ideal as possible. This is usually not a problem to obtain in image processing, but often difficult to achieve within the field of seismic processing. For example, no ideally unblended data exist that can serve as the ground truth when training on blended-by-acquisition (real field blended) data. Similarly, it is not feasible to acquire seismic data containing primaries only, for the training of demultiple through a supervised learning approach.

In view of this situation, one choice is the use of results output from a conventional deblending flow as ground truth. However, preserving the weak useful signals while removing the strong overlapping blending noise is a challenging task which means that no existing deblending algorithm has managed to produce deblended data that are as perfect as unblended. Note again that in this study the term 'unblended shot gather' only refers to the shot gather acquired with no overlapping shot followed. However, in the literature the same terminology may also refer to the deblended shot gather output from a deblending algorithm.

An alternative approach is to employ synthetic data as the ground truth in the training process. However, this also represents a major challenge due to the complexity of field data. Thus, models trained on synthetic data need to be retrained or fine-tuned on a selected part of the field data before being applied on the remaining part (Li et al., 2020). A possible improvement has been proposed by Wang et al. (2020), where the synthetic training data are improved iteratively from velocity-model updates based on deblending predictions from a CNN employing the field acquisition parameters. No matter which of these two approaches are used, the ground truth is not obtained directly from the survey of interest and additional time and labor costs beyond the blended acquisition are required.

To overcome the fundamental challenges described above, new strategies to generate training data have been proposed during recent years. In a very recent study on ML-based deblending of onshore Distributed Source Array (DSA) field data, Baardman et al. (2020) proposed to use an earlier acquired unblended data set from the vicinity of the DSA acquisition to generate the training pairs and demonstrated its feasibility. However, such a strategy still has limitations in new exploration areas where little or no relevant data exists.

In this paper, we propose a novel strategy to generate training data suitable for any type of blended acquisition. In blended surveys, and independent of the actual acquisition design, the later samples in the last shot gathers from all streamers in different sail lines contain no blending noise and with long enough recording length as shown in Figure 3b. By utilizing these shots we ensure that high-quality ground truth data are always available without extra acquisition cost. Since the unblended last shot gathers are obtained directly as part of the survey, accessing them requires no additional time or labor costs, such as running a time-consuming conventional deblending algorithm or forward modeling. Neither does this strategy rely on the existence of conventionally acquired seismic data from the vicinity of the blended survey. By manually blending shot gathers from the last shootings, we can obtain high-quality training pairs with good control of the ground truth and fully adapted to the given survey area.

Take the source-over-streamer blended acquisition discussed in the previous section as an example. In this acquisition, 16 streamers and a total of 160 sail lines were employed. The unblended shot gathers were selected from 14 streamers of 146 sail lines but excluding those that were contaminated by the Front-Source or had missing traces. From this set of data, we constructed blended shot gathers by adding two unblended shot gathers with a fixed delay time perturbed with a predefined random jitter. By changing the delay time, various source configurations can be simulated. In the example considered, the delay time was set to 3.0 s (same as the shooting rate in the field survey of interest) ± 0.5 s of random jitter. Employing this set of data, we were in the position to compare the performance of our DNN in case the ground truths were respectively deblended results from a conventional algorithm and the unblended last shot gathers of a blended acquisition. The training, validation, and test data sets consisted of 7000, 1000, and 200 images, respectively, with 256 traces per image.

The conventional deblending algorithm employed in this study is based on an advanced workflow used in industry-scale production. It consists of a L1 inversion-based 3D deblending method (Peng and Meng, 2016) for towed-streamer data combined with a post-processing workflow. The L1 inversion-based method pursues a sparse representation of 3D coherent signals that matches the blended data in a domain based on a combination of 2D Tau-P and 2D directional wavelet transforms. A sparse 2D Tau-P transform (Trad et al., 2003; Peng and Meng, 2016) was first applied to the shot gather to focus coherent signals along channels to corresponding constant-P traces. Subsequently, an L1 inversion algorithm based on the 2D high-angular resolution complex wavelet transform (HARCWT) (Peng et al., 2013) was employed to deblend the signals of different sources for each common-P gather. To further improve the deblending result, the output from the inversion-based algorithm containing blending noise residuals was fed into a post-processing workflow that mainly performs a residual denoising.

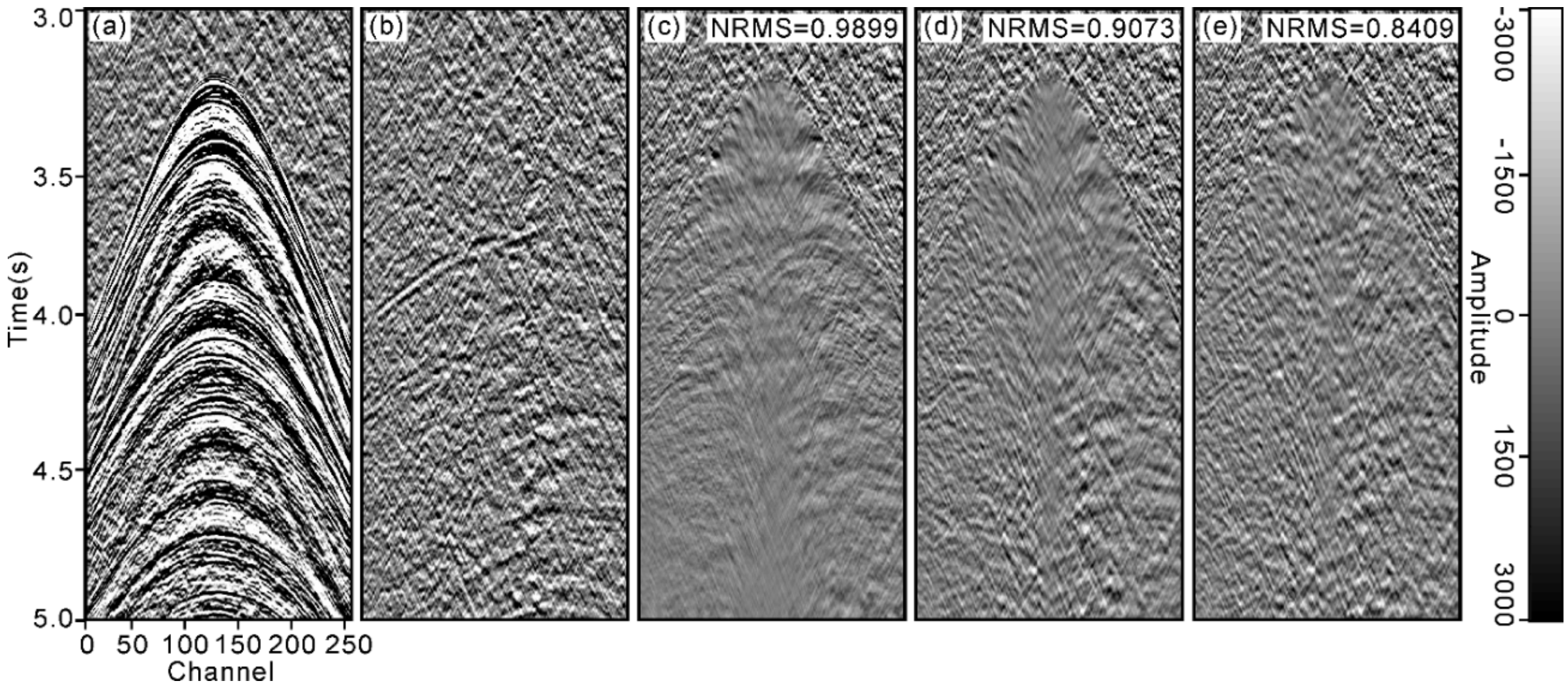

Figure 6: (a) Blended data, (b) unblended data, (c) deblending result from the employed DNN when the training pairs are generated using deblending results from a conventional algorithm as the ground truth, (d) deblending result from the same network when the training pairs are generated using the unblended last shot gathers from the blended acquisition as the ground truth and (e) deblending result from the same network when adding two adjacent blended shot gathers

(one on each side) in the input of (d) as additional channels.

Figures 6a and 6b show respectively the (manually) blended and unblended shot gather. The root mean square (RMS) amplitude of the blended trace with the longest record of blending noise in Figure 6a is 26326 while the RMS amplitude of the corresponding unblended trace in Figure 6b is only 1507. Figures 6c and 6d show the deblending performance of our network when the ground truths are generated by either of the two strategies: deblending results obtained from use of the conventional algorithm or unblended last shot gathers from the blended acquisition. Compared to the ground truth in Figure 6b, we observe that both Figure 6c and 6d suffer from some amplitude dimming. This is slightly worse after 4.5 s for Figure 6c, which compounds dimming from the conventional algorithm with dimming for the network-based approach. Processing quality of future data by a trained network is based on the validity of its previous example-data experience. Thus, the training data determines somehow the upper limit of the network performance.

To quantify the quality of deblending, we calculated one commonly used measurement of repeatability, the normalized root mean square (NRMS) amplitude of the difference between unblended and deblended traces. The RMS amplitude $x_{w,t}$ of the $w^{\text{th}}$ trace of a shot gather defined within a time window limited by traveltimes $t_1$ and $t_2$ is given as

$$\boldsymbol{RMS}(x_{w,t}) = \sqrt{\frac{\sum_{t=t_1}^{t=t_2}(x_{w,t})^2}{N_{t_1-t_2}}}, \tag{2}$$

where $N_{t_1-t_2}$ is the number of samples in the time interval $t_1 - t_2$. The NRMS amplitude of the difference between the unblended trace $u_{w,t}$ and the deblended trace $d_{w,t}$ is expressed as

$$\boldsymbol{NRMS}(u_{w,t}, d_{w,t}) = \frac{2 \times \boldsymbol{RMS}(u_{w,t} - d_{w,t})}{\boldsymbol{RMS}(u_{w,t}) + \boldsymbol{RMS}(d_{w,t})}. \tag{3}$$

The values of the NRMS are limited to the range 0 to 2. If a phase shift of 180 degrees exists between the two traces or if one trace contains only zeros, the NRMS equals to 2. The lower the NRMS, the better repeatability between the traces. By taking the mean of the NRMS of all $W$ traces in the shot gather, we obtain a measure on the similarity between the unblended and deblended shot gathers

$$\boldsymbol{NRMS} = \frac{1}{W}\sum_{w=w_1}^{w=w_2}\left(\boldsymbol{NRMS}\left(u_{w,t}, d_{w,t}\right)\right). \tag{4}$$

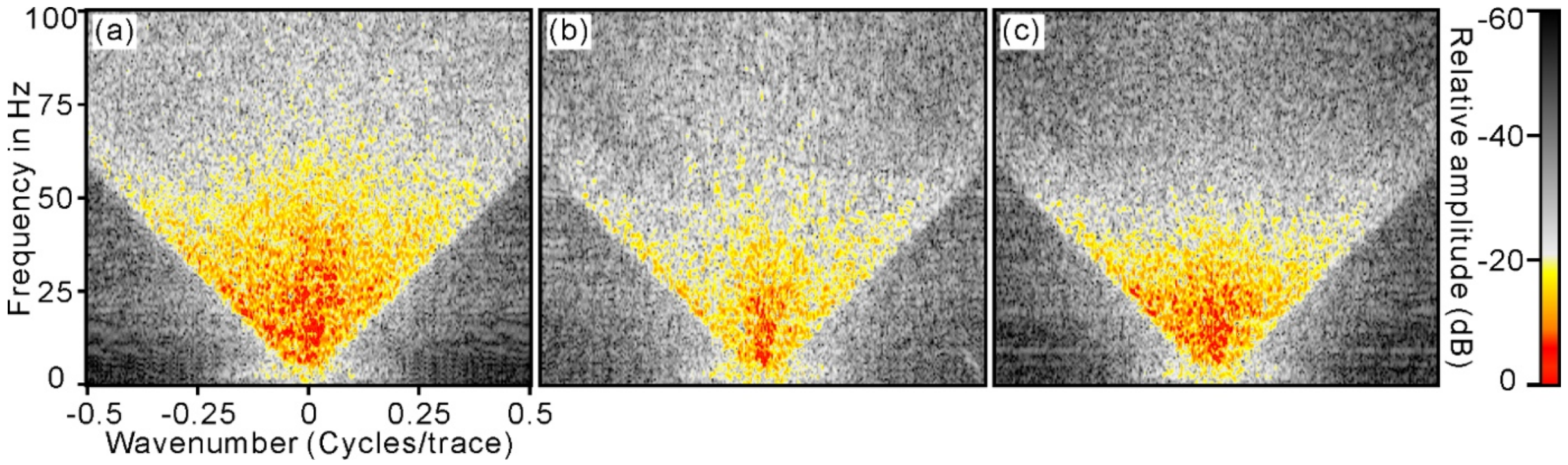

Figure 7: Frequency-wavenumber spectra of (a) unblended data, (b) deblending result from the employed DNN when the training pairs are generated using deblending results from a conventional algorithm as the ground truth and (c) deblending result from the same network when the training pairs are generated using the unblended last shot gathers from the blended acquisition as the ground truth.

In the case of training pairs generated by our proposed strategy (Figure 6d, single-channel input), the NRMS value of the U-Net deblended shot gather considered is about 0.91 which is less than the corresponding value (about 0.99) in case of using deblending results from a conventional algorithm as the ground truth (Figure 6c). These observations are also supported by the frequency-wavenumber spectra as shown in Figure 7. Figure 7a shows the frequency-wavenumber spectrum of the unblended data (Figure 6b). Figures 7b and 7c show the frequency-wavenumber spectra of the deblending results from the same employed DNN but with two different strategies to generate the training pairs. Compared to Figure 7b where deblending results from a conventional algorithm are used as the ground truth, Figure 7c, where the unblended last shot gathers from the blended acquisition are used as the ground truth, is obviously closer with Figure 7a with significantly less signal leakages.

## DATA CONDITIONING

To further improve the deblending quality of the DNN, we introduced a series of pre-conditioning steps that can be selectively used:

1. Add adjacent blended shot gathers in the input as additional channels.
2. Train the network to predict both primary-source events and the down-scaled blending noise.
3. Align the to-be-deblended shot gathers by blending noise instead of by primary-source events.

### Use of adjacent shot gathers as additional channels of the input

In general, images can be represented by third-order tensors, characterized by height, width, and the number of channel(s). The height and width of an image relates spatial information whereas the concept of channels assigns a multi-dimensional representation to each pixel location. As an example, digital color images are represented by three standard channels (RGB channels) reflecting the amount of those three primary colors. 2D seismic data can be regarded as grayscale images with a single channel only. Thus, a simple regression task involving 2D seismic data allows the use of 2D tensors to represent input, filter kernels, and output. In DNN-based deblending, the mapping from a blended shot gather to the predicted primary-source events can be represented as

$$\boldsymbol{NET}(\boldsymbol{S}_{i,k}^{j}) \rightarrow \boldsymbol{P}, \tag{5}$$

where $\boldsymbol{NET}$ represents the network-based approach, $\boldsymbol{S}_{i,k}^{j}$ represents the blended shot gather whose primary-source gather is fired by the $k^{\text{th}}$ shooting of $\boldsymbol{S}_i$ and received by the $j^{\text{th}}$ streamer. Moreover, $\boldsymbol{P}$ represents the predicted primary-source events of $\boldsymbol{S}_{i,k}^{j}$.

Seismic events of consecutive shot gathers are obviously correlated, with the degree of correlation determined by the number of shots selected and the complexity of the geology. We therefore propose to train the network employing multi-channel input data that includes adjacent blended shot gathers on both sides of the to-be-deblended shot gather as additional channels. In each sample, two groups of adjacent blended shot gathers with different alignment can be alternatively included. In one group, the adjacent shot gathers are aligned with the to-be-deblended shot gather employing the primary-source events. In the other group, the alignment is based on the blending noise. When adding more channels, both inputs and the hidden representations become 3D tensors. Using a CNN to map a multi-channel input consisting of multiple blended shot gathers to the primary-source events in the to-be-deblended shot gather can be written as

$$\boldsymbol{NET}(\boldsymbol{S}_{i,k-n}^{j}, \dots, \boldsymbol{S}_{i,k-1}^{j}, \boldsymbol{S}_{i,k}^{j}, \boldsymbol{S}_{i,k+1}^{j}, \dots, \boldsymbol{S}_{i,k+n}^{j}) \rightarrow \boldsymbol{P}, \tag{6}$$

where $\boldsymbol{S}_{i,k}^{j}$ is the to-be-deblended shot gather. Correspondingly, $\boldsymbol{S}_{i,k-n}^{j}, \dots, \boldsymbol{S}_{i,k-1}^{j}$ represent $n$ adjacent blended shot gathers fired earlier than $\boldsymbol{S}_{i,k}^{j}$. In the same way, $\boldsymbol{S}_{i,k+1}^{j}, \dots, \boldsymbol{S}_{i,k+n}^{j}$ represent $n$ adjacent blended shot gathers fired later than $\boldsymbol{S}_{i,k}^{j}$. Finally, $\boldsymbol{P}$ denotes the predicted primary-source events of $\boldsymbol{S}_{i,k}^{j}$.

Note that in a field blended acquisition, Equation 6 is only suitable for the preparation of the test data set. This is because no consecutive unblended shot gathers exist that can be blended into multi-channel input data for the training process. Therefore, we propose to use adjacent streamers from the last shot gather in each sail line as the channels of the training sample. The training process can then be represented as

$$\boldsymbol{NET}(\boldsymbol{S}_{K}^{j-n}, \dots, \boldsymbol{S}_{K}^{j-1}, \boldsymbol{S}_{K}^{j}, \boldsymbol{S}_{K}^{j+1}, \dots, \boldsymbol{S}_{K}^{j+n}) \rightarrow \boldsymbol{P}, \tag{7}$$

where $\boldsymbol{S}_K^j$ is the to-be-deblended manually blended shot gather whose primary-source gather originates from the last shooting in a given sail line and received by the $j$th streamer. In Equation 7, $\boldsymbol{S}_K^{j-n}, \ldots, \boldsymbol{S}_K^{j-1}$ and $\boldsymbol{S}_K^{j+1}, \ldots, \boldsymbol{S}_K^{j+n}$ represent the manually blended shot gathers whose primary-source gathers originate from the last shooting of the same sail line and received by the *n* adjacent streamers on both sides of the $j$th streamer. Finally, $\boldsymbol{P}$ denotes the predicted primary-source events of $\boldsymbol{S}_K^j$.

We have already discussed the result shown in Figure 6d representing deblending results obtained employing the DNN with a single-channel input only. Correspondingly, the deblending results obtained employing the multi-channel approach as discussed above are shown in Figure 6e. In this case, the input data contains three channels which are the to-be-deblended shot gather and one adjacent blended shot gather on each side. The performance of our network has now improved with less dimming of primary-source events in combination with an efficient attenuation of blending noise. The NRMS value has further decreased for this example shot gather from 0.91 (single-channel input) to 0.84 (multi-channel input).

**Predicting both primary-source events and the down-scaled blending noise**

When the network is trained to focus on the single task of predicting the primary-source events only, information about the blending noise that might help it perform even better is ignored. For such a case, the loss function of our network model employing the mean square error (MSE) can be written as

$$\boldsymbol{Loss} = \boldsymbol{MSE}(\boldsymbol{P},\ \widehat{\boldsymbol{P}}), \tag{8}$$

where $\boldsymbol{P}$ is the predicted primary-source events and $\widehat{\boldsymbol{P}}$ is the corresponding ground truth.

In order to allow the network to make a more comprehensive use of the training data, we apply the multi-task learning (MTL) technique (Ruder, 2017) which enables the model to generalize better on the original task by sharing representations between related tasks. Due to an extremely poor SNR in the blended section (the blending nose is typically two orders of magnitude higher than the underlying signal), it was very difficult for the network to separate weak signals of the primary-source from the overlapping blending noise. This is especially true with data in the shot domain where these events share many similarities. Use of MTL can therefore help the network to train in a more robust manner. Thus, with the main task being the prediction of the primary-source events, we add in the prediction of blending noise as a related and auxiliary task.

Note that the amplitude of the blending noise $\boldsymbol{N}$ is much higher than the primary energy $\boldsymbol{P}$ in the area of interest (cf. Figures 2, 3 and 4), therefore the loss function is biased toward the task of predicting the blending noise $\boldsymbol{N}$. In order to balance the terms of primary-source events and blending noise, we propose to scale down the blending noise by a factor $\boldsymbol{\lambda}$ during the training and validation process. Thus, the modified loss function of our network model takes the form

$$\boldsymbol{Loss} = \boldsymbol{MSE}([\boldsymbol{P},\ \boldsymbol{\lambda} * \boldsymbol{N}],\ [\widehat{\boldsymbol{P}},\ \boldsymbol{\lambda} * \widehat{\boldsymbol{N}}]), \tag{9}$$

where $\boldsymbol{\lambda} * \boldsymbol{N}$ represents the predicted blending noise with down-scaling applied when $\boldsymbol{\lambda} < 1$, which forms a 2-channel output $[\boldsymbol{P},\ \boldsymbol{\lambda} * \boldsymbol{N}]$ with the predicted primary-source events $\boldsymbol{P}$. Correspondingly, $\widehat{\boldsymbol{N}}$ represents the true blending noise and it forms the 2-channel ground truth $[\widehat{\boldsymbol{P}},\ \boldsymbol{\lambda} * \widehat{\boldsymbol{N}}]$ with the true primary-source events $\widehat{\boldsymbol{P}}$. The choice $\boldsymbol{\lambda} = 1$ implies that no scaling is applied to the blending noise.

**Align input data based on blending noise**

In field blended data, blending noise appears on consecutive shot gathers at different times due to the use of random jitters. In the applications of conventional deblending algorithms, blended shot gathers are always aligned by the primary-source events as shown in Figure 8a. Thus, after being resorted from the common source to common offset or common receiver domain, the primary-source events of the blended data will exhibit a continuous and coherent form but the blending noise will manifest itself as incoherent contributions (Mahdad et al., 2011).

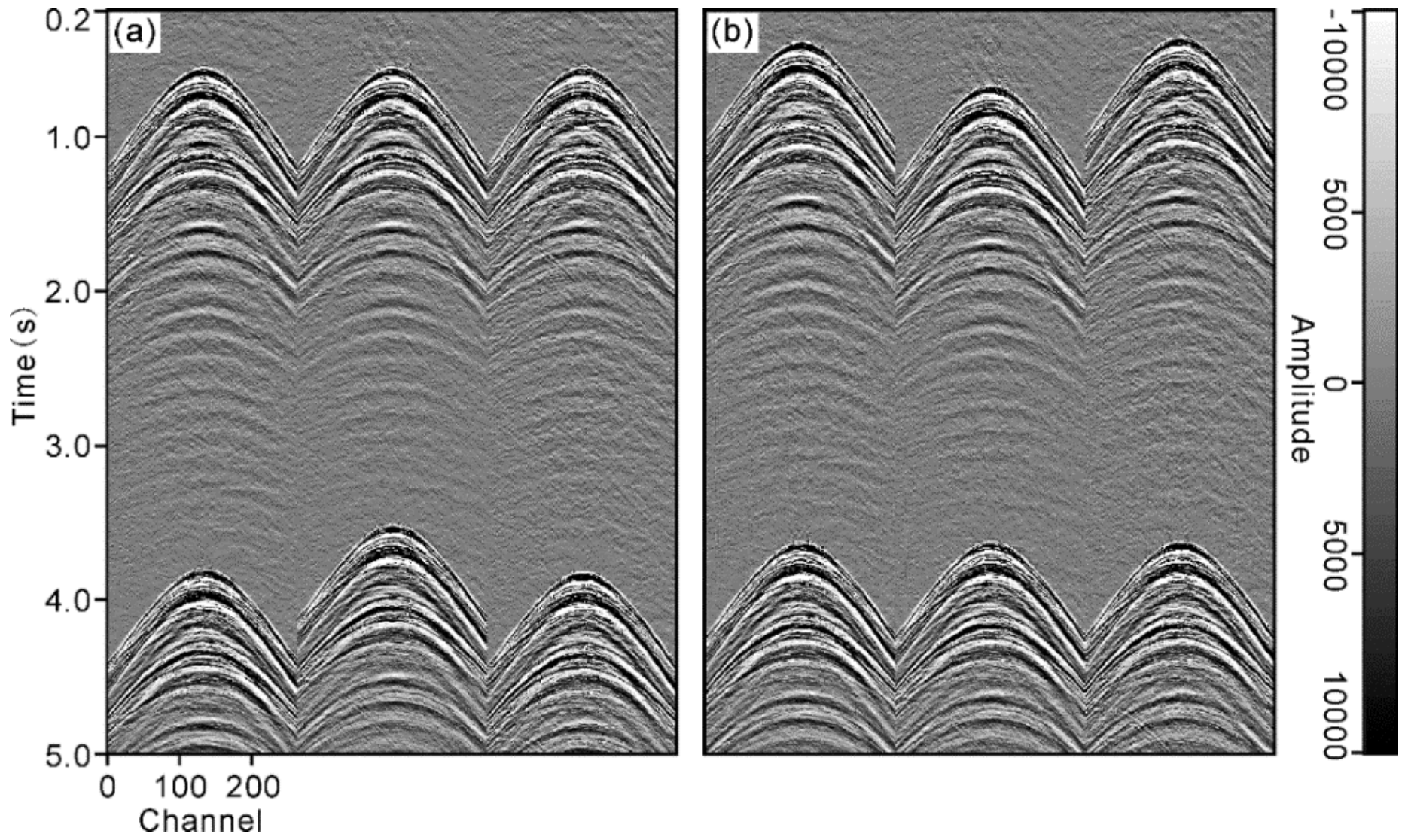

Figure 8: Examples of alignment based on (a) primary-source events or (b) blending noise of the to-be-deblended shot gathers.

Our proposed DL approach performs the predictions directly in the shot domain where both the primary-source events and the blending noise are continuous and coherent events. Aligning the to-be-deblended shot gathers by the primary-source events leads to a training data set having a relatively high variance since the starting time of the blending noise varies among the samples. Such a case generally requires a more complex network model to train on. Therefore, we propose as an alternative to align the to-be-deblended shot gathers by the blending noise as shown in Figure 8b. Tests on field data shows that this type of alignment improves the performance of our network.

## FIELD DATA DEMONSTRATION

To further test the quality and performance of our DL approach in a rigorous manner, we applied it to real blended-by-acquisition data. The field data used in this section are taken from the same source-over-streamer blended acquisition as discussed before. As a *benchmark method*, we apply the same conventional workflow introduced earlier. All comparisons are carried out in the stacked domain. Training, validation and testing of the network are all performed in the shot domain.

We also introduce a so-called *reference case* for our network with no data conditioning steps applied. Thus, in the reference case both input and output of the network are single-channeled. Further, the blended shot gathers are aligned by the primary-source events, and the network is trained to perform a prediction of the primary-source events only.

Correspondingly, the *proposed approach* implies use of the network with all data conditioning steps included. The number of adjacent shot gathers to be used on each side of the to-be-deblended shot gather was first tested. The optimal number was selected based on the minimum validation loss during the training process and the corresponding training time. The training, validation, and test data sets consisted of 7000, 1000 and 200 images, respectively. As shown in Figure 9, the use of 2 adjacent shot gathers on each side was found to perform best in this study. Moreover, we employed an alignment based on the blending noise and a combined prediction of both the primary-source events and the down-scaled blending noise ($\lambda$ was set to 0.04). Note that the two predictions from the DNN can alternatively both be used in further processing. In this study, we employed the predicted primary-source events only.

The use of a multi-channel input will make the DNN more computationally demanding at the training stage because of the increased input data volume. Figure 10 shows the losses on the training and validation data sets of both the reference case and the proposed approach. The proposed approach took around 5.6 minutes to train one epoch compared to the reference case consuming around 4.7 minutes on Nvidia Quadro RTX 6000 GPU. The complete training process of the proposed approach took only around 5.5 hours in this study. It should also be noticed that once finished the training, both the reference case and the proposed approach took less than 0.3 s to deblend a single shot gather at the application stage. The conditioning step of using multi-channel inputs barely changed the application cost.

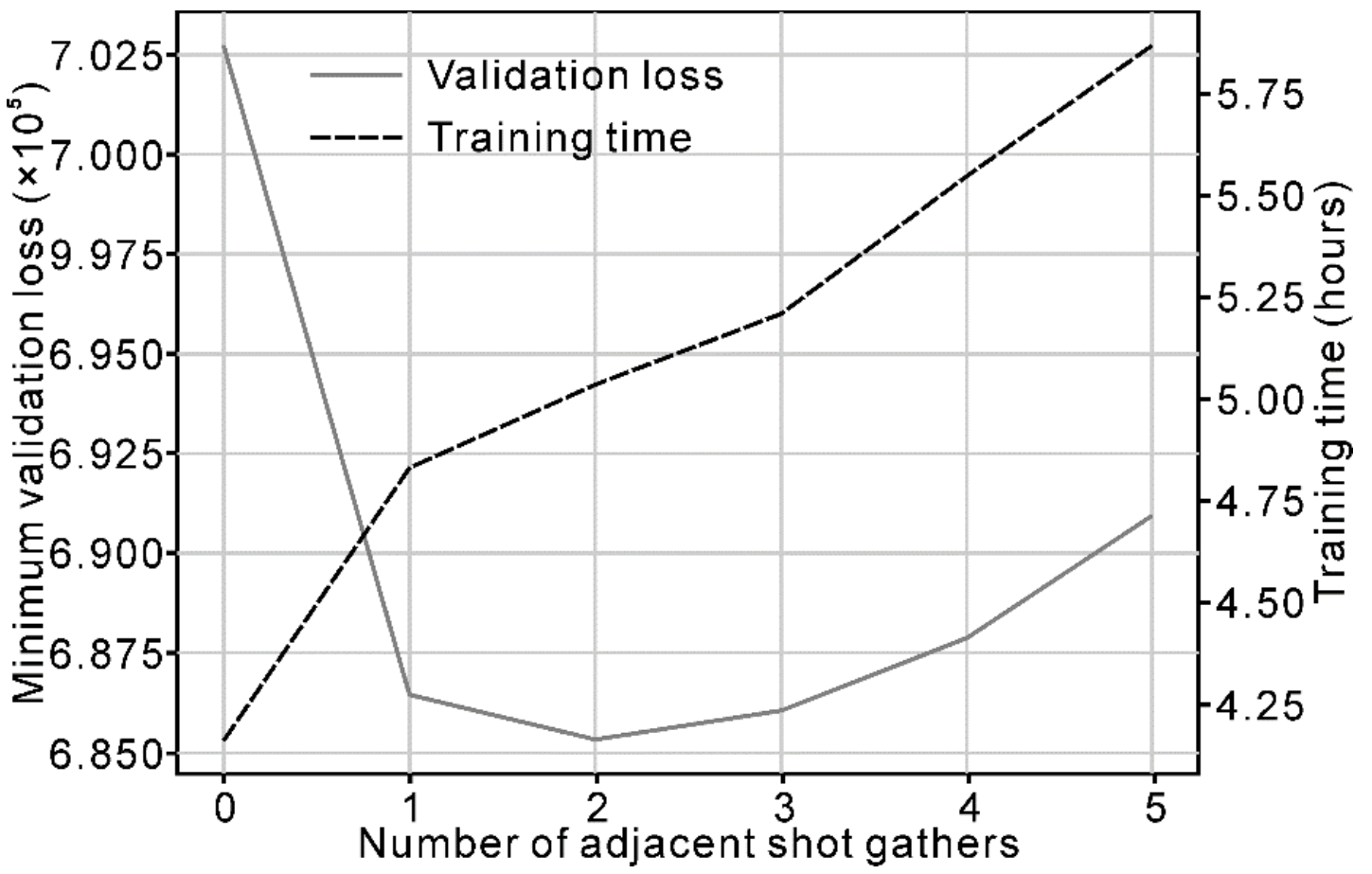

Figure 9: Variation of final validation loss and training time when changing the number of adjacent shot gathers.

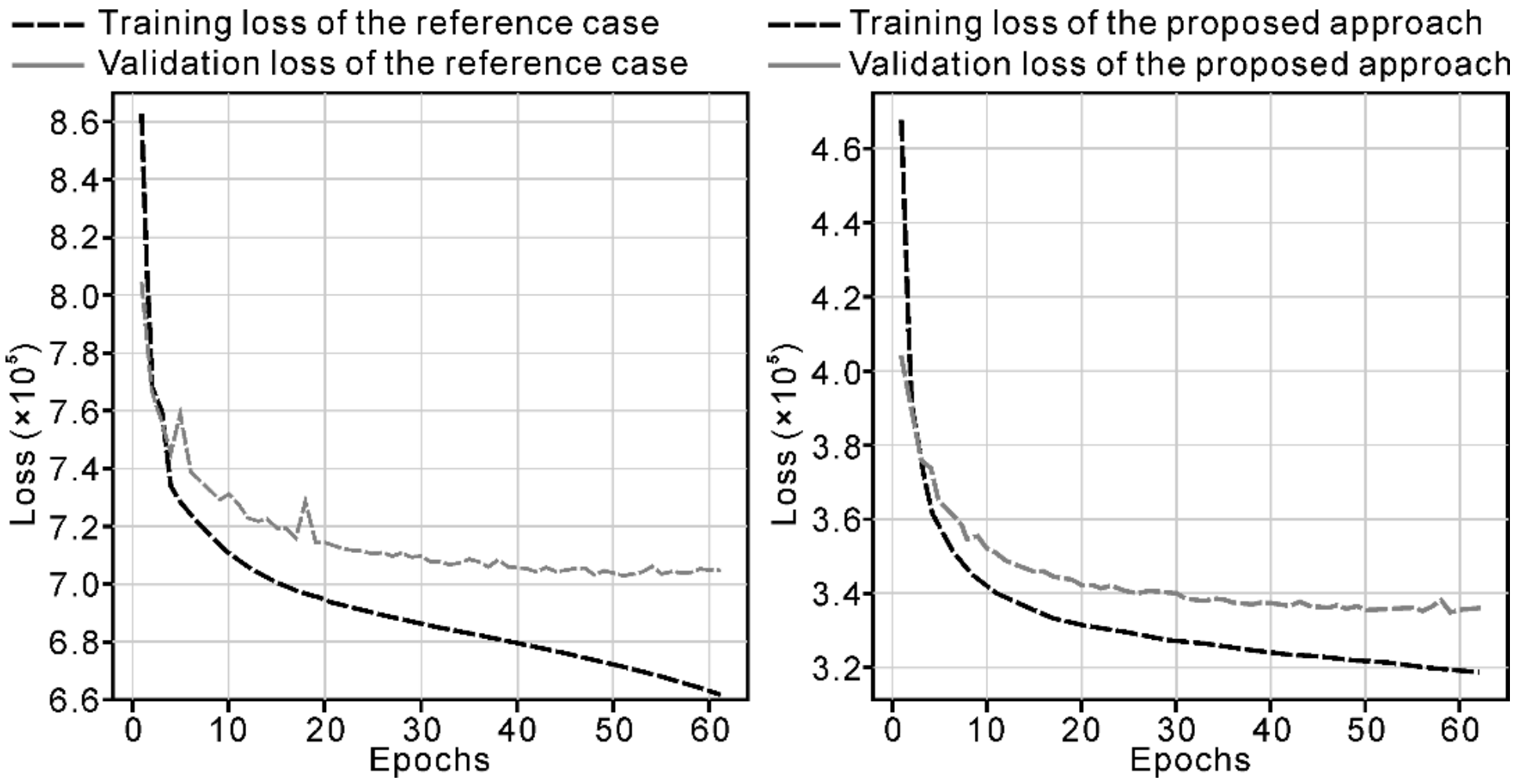

Figure 10: Training and validation losses of the reference case and the proposed approach.

Figure 11 shows a CMP stack of the blended-by-acquisition data from the Castberg field in the Barents Sea (Vinje and Elboth, 2019; Poole et al., 2020). About 4000 shots from one sail line were used. The corresponding deblending results of the reference case after CMP- stacking is shown in Figure 12. Correspondingly, Figure 13 shows the deblending result obtained in the CMP-stacked domain if the proposed approach is applied. Finally, the deblending results obtained using the conventional workflow is shown in Figure 14. Note that all CMP stacks are displayed from 3.0 to 5.0 s two-way traveltime (TWT) to focus only on the target area.

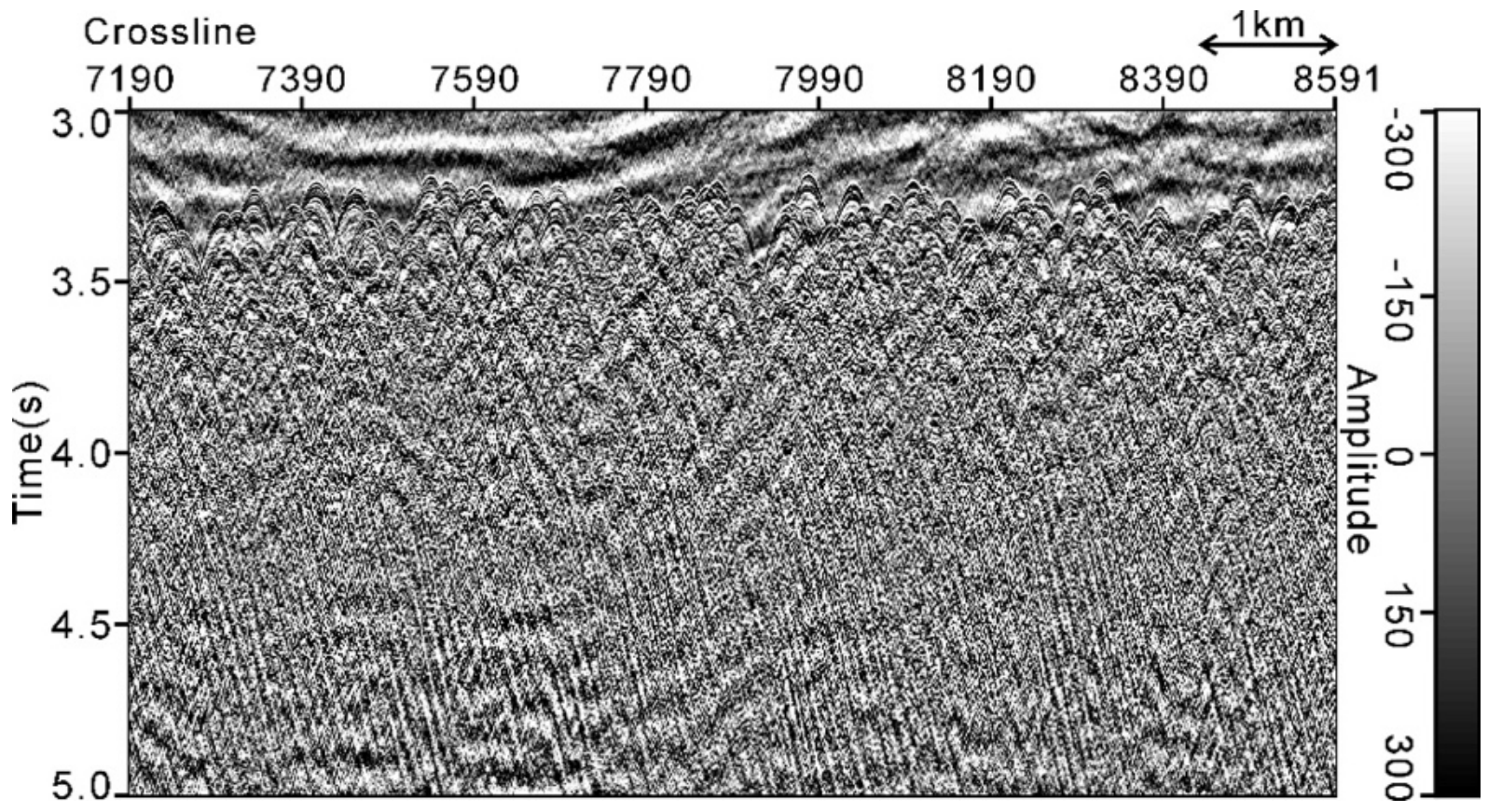

Figure 11: CMP stack of the blended-by-acquisition data from the Barents Sea.

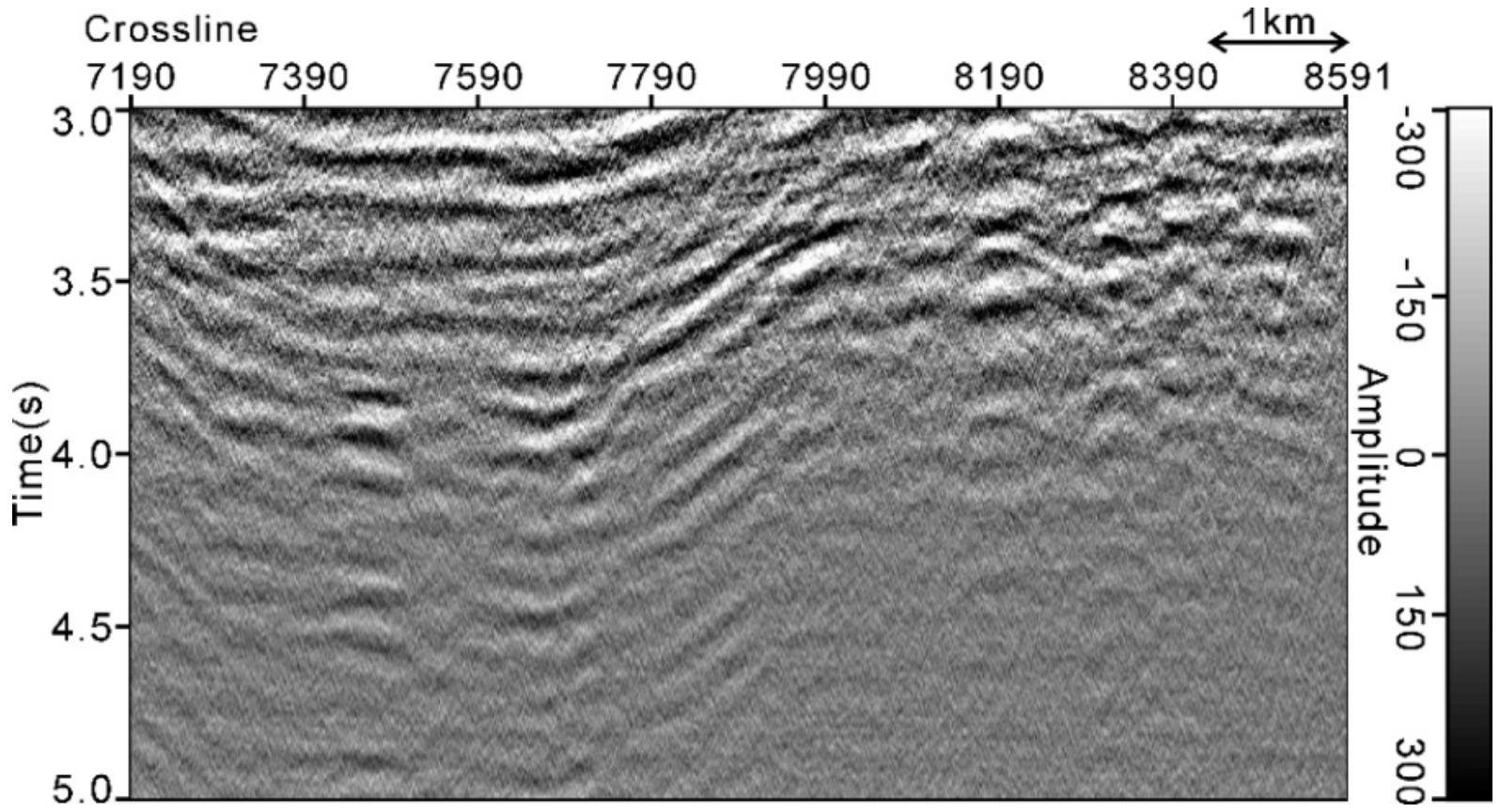

Figure 12: Deblended stack of the reference case (with no data conditioning).

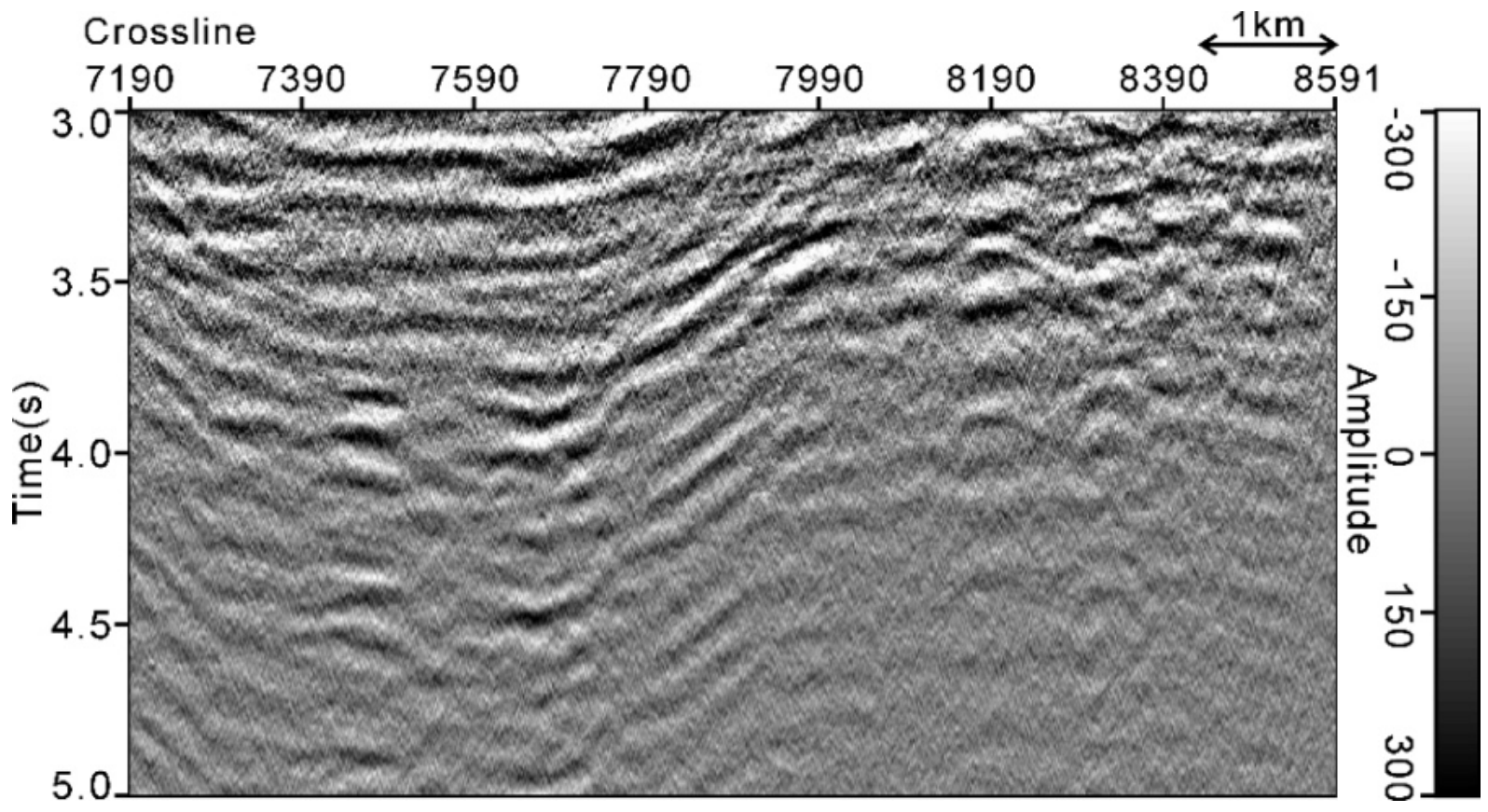

Figure 13: Deblended stack of the proposed approach (with complete data conditioning).

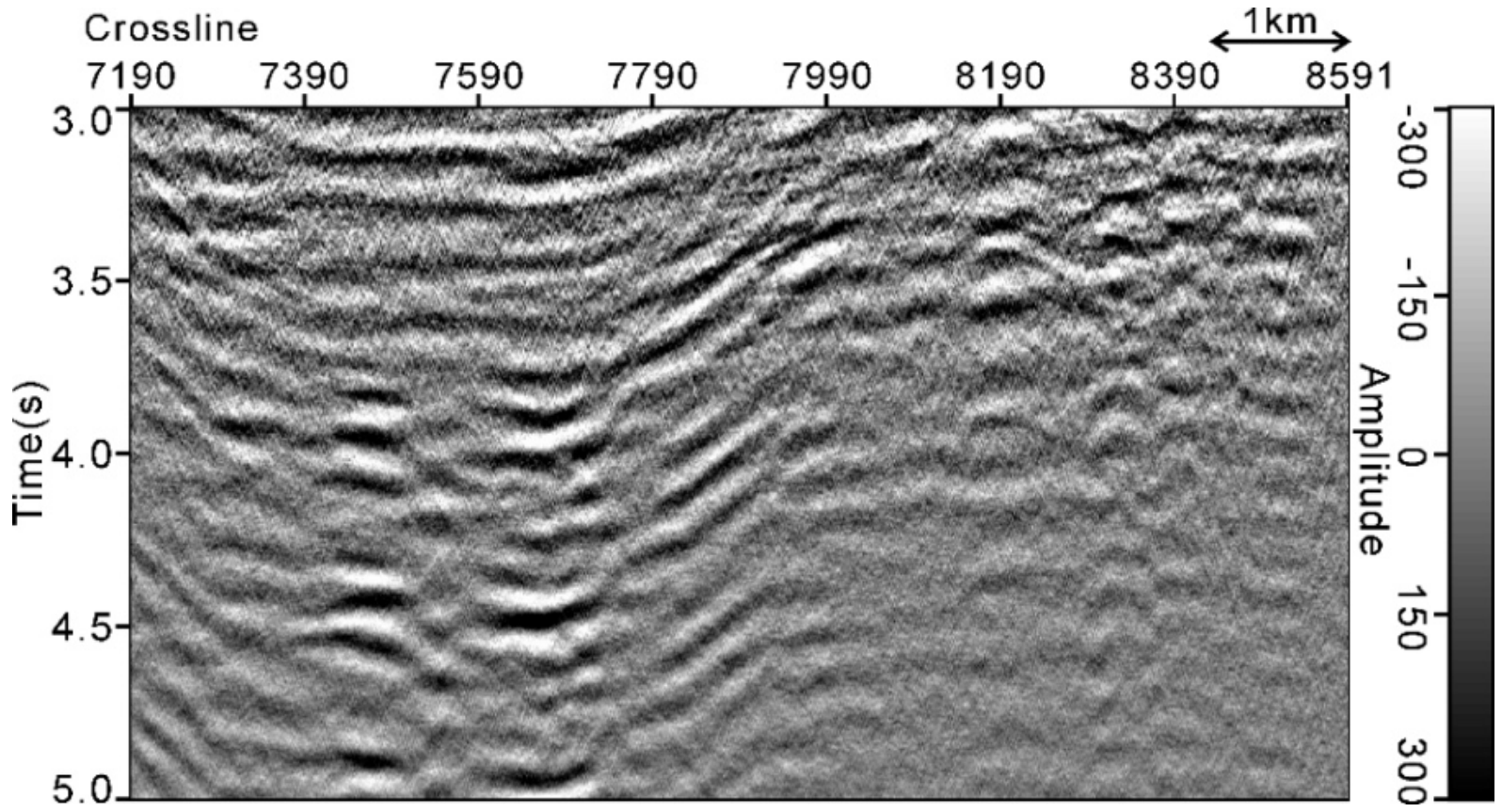

Figure 14: Deblended stack of a conventional algorithm.

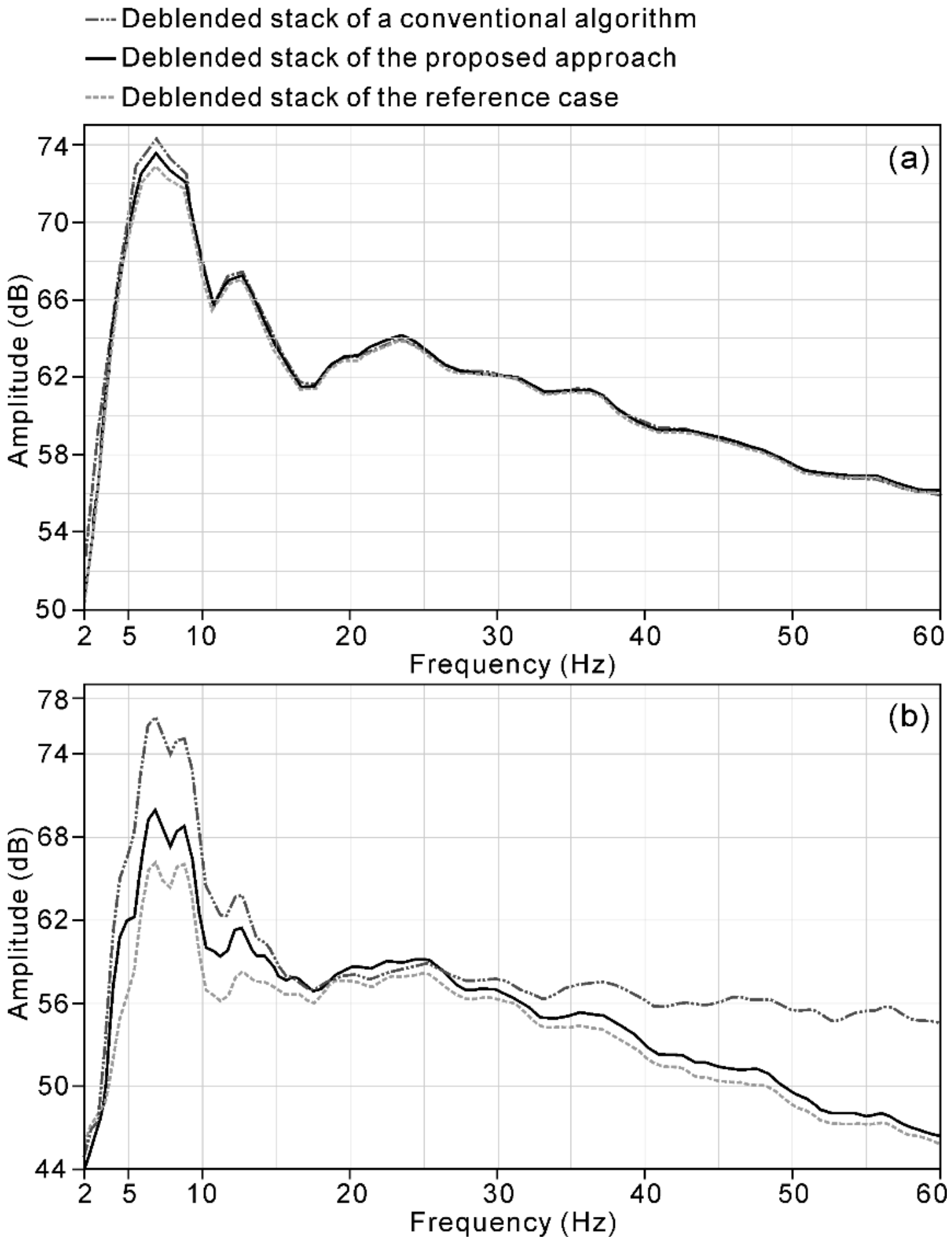

Figure 15: Frequency spectra of the deblended stacks obtained from a conventional algorithm, the reference case and the proposed approach within the time interval (a) from 3.2 to 4.0 s TWT and (b) from 4.0 to 5.0 s TWT.

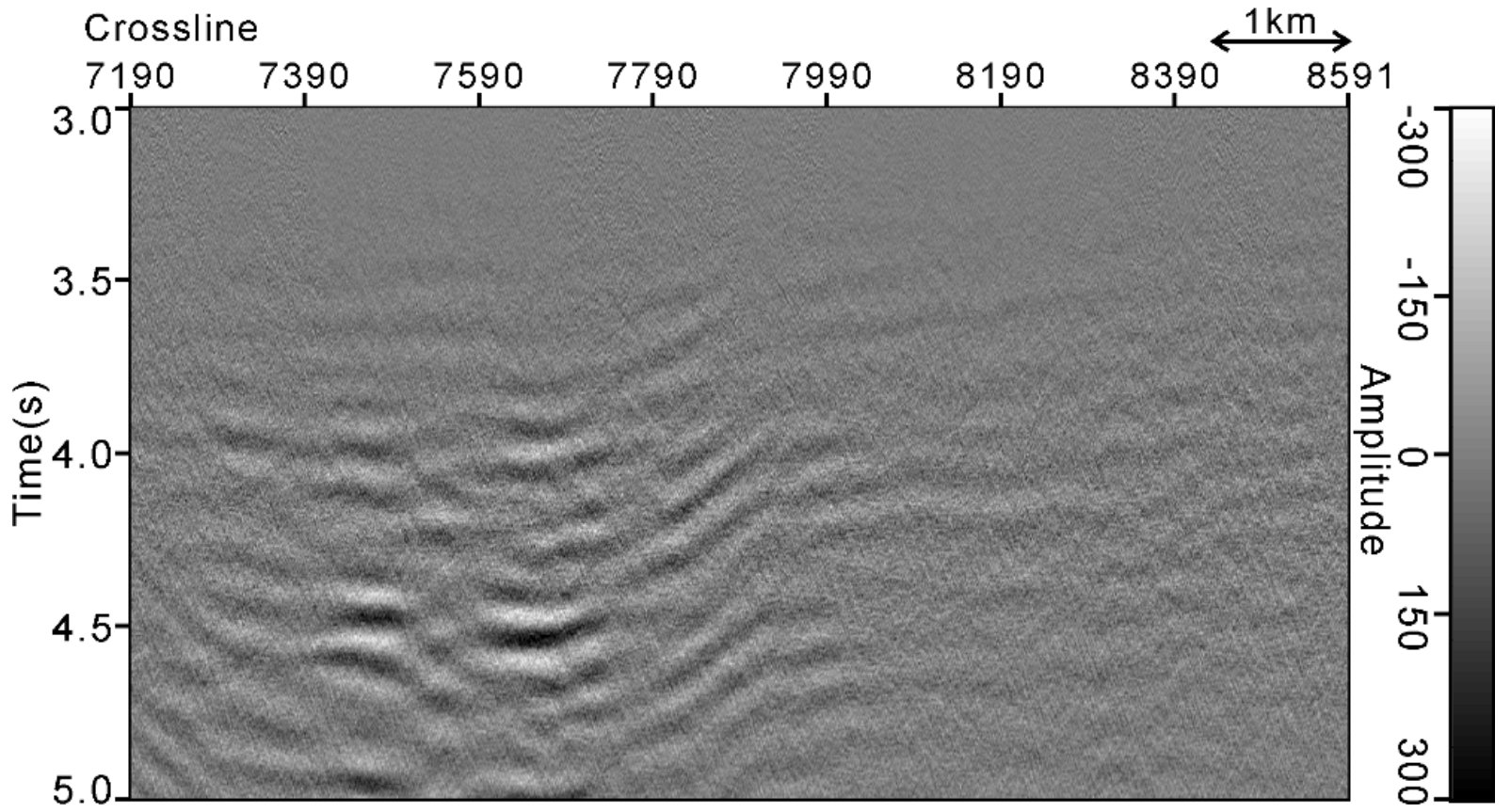

Figure 16: Difference between deblended stacks of the proposed approach and the conventional algorithm.

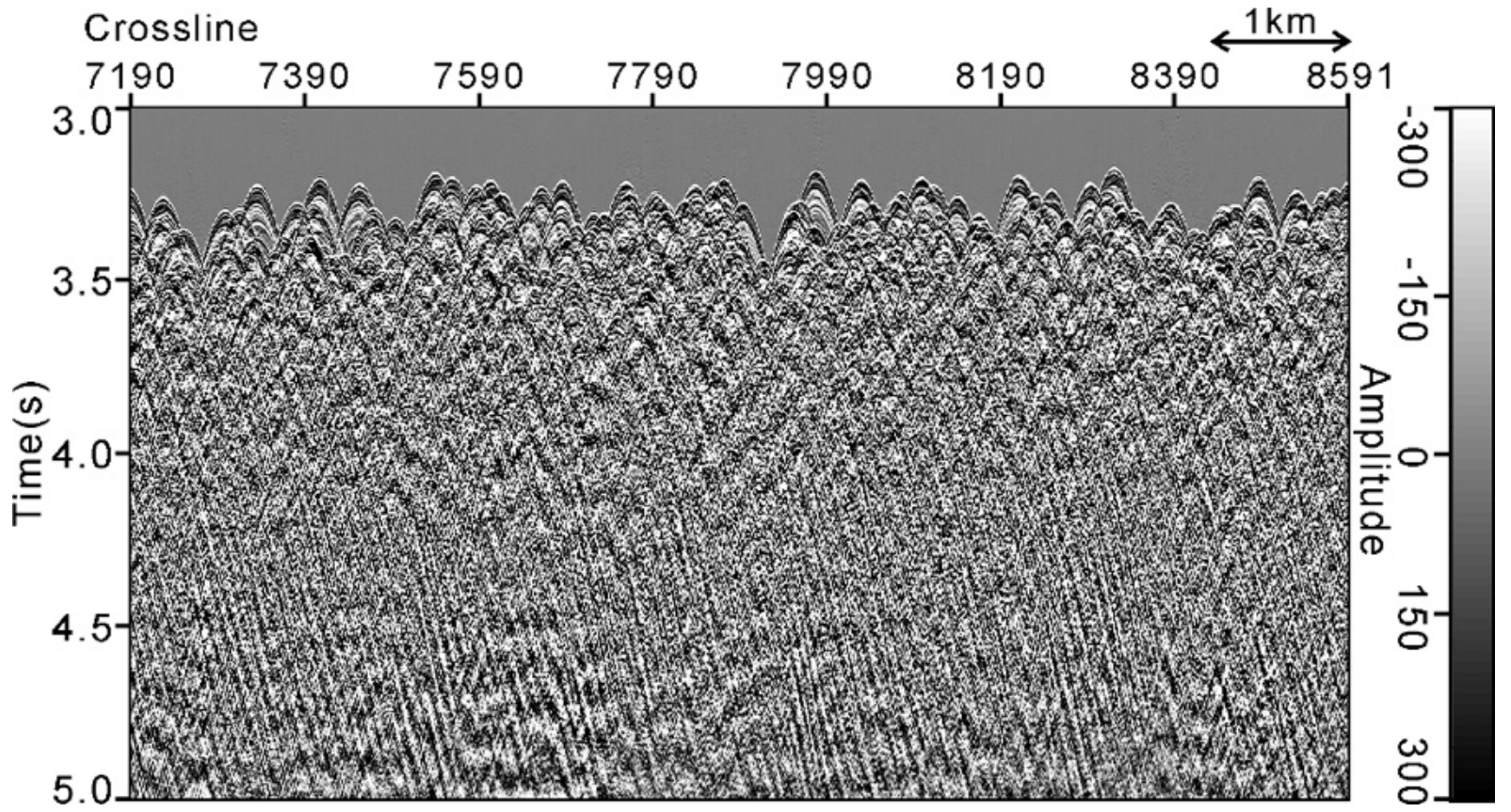

Figure 17: Blending noise removed by the proposed approach.

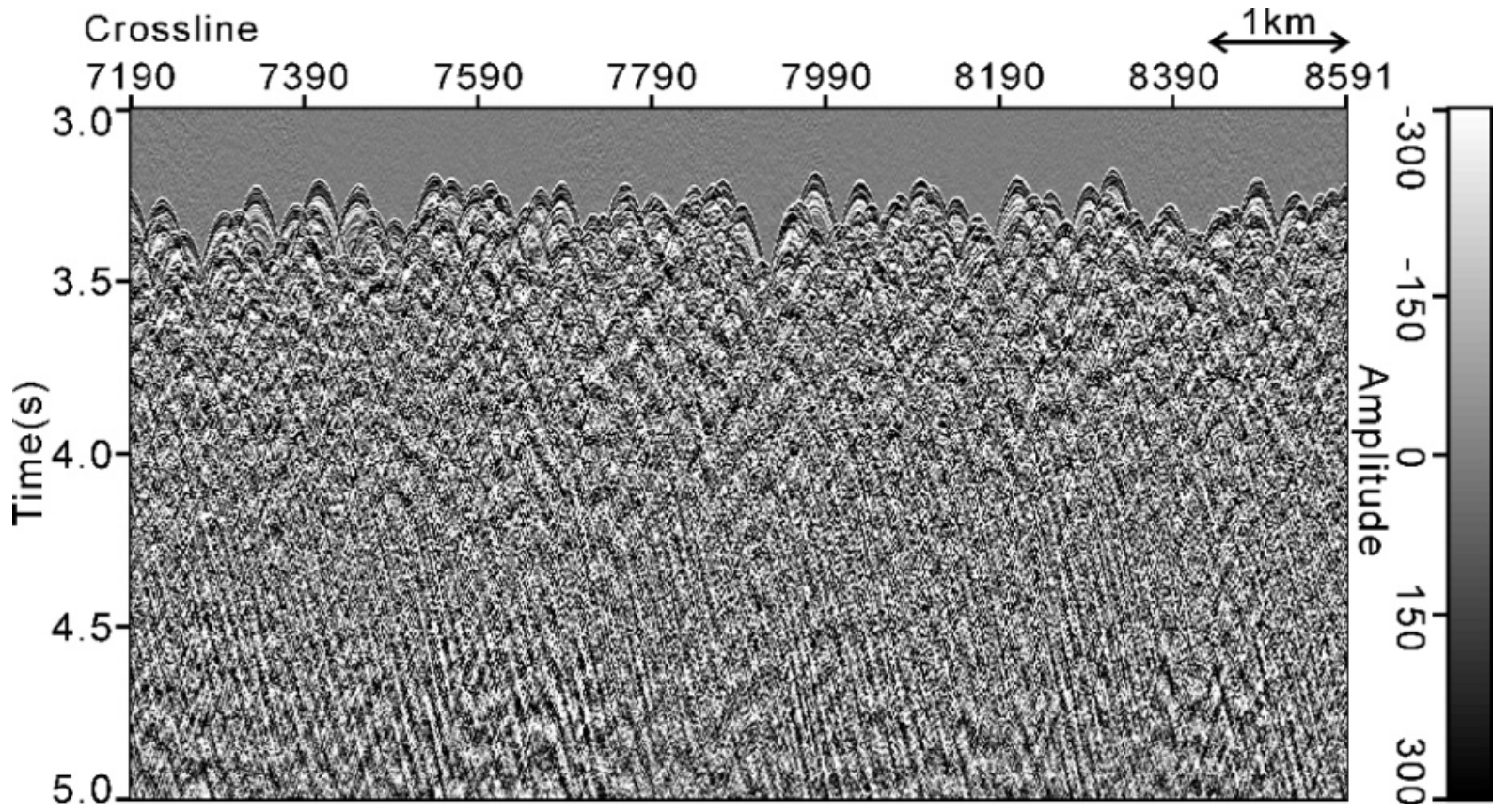

Figure 18: Blending noise removed by a conventional algorithm.

As can be seen from Figure 11, the blending noise starts to appear at around 3.2 s TWT

with very strong energy which masks out primary shot energy. Within the time interval from 3.2 to 4.0 s TWT, the proposed approach (Figure 13) and also the reference case (Figure 12) perform very well with deblending results being almost as good as in the case of the conventional algorithm (Figure 14). These observations are also supported by the frequency spectra of the deblended stacks computed within the same time interval as shown in Figure 15a. Corresponding frequency spectra for the time interval between 4.0 and 5.0 s TWT are shown in Figure 15b. Within this time interval, the conventional workflow performs best, but the proposed approach gives also an overall satisfactory result, where the blending noise has been efficiently removed but with slightly dimmer primary-source events and a loss of high frequencies. Direct comparison between Figures 12 and 13 also demonstrate that with data conditioning included the proposed approach can significantly improve the deblending effect of the employed DNN compared to the reference case although some primary-source energy leakage is still observed in the deep section as can be observed in Figure 16. For completeness, we plotted the stacked blending noise removed by the proposed approach and the conventional algorithm in Figures 17 and 18, respectively. Blending noise removed by these two approaches are similar in the CMP-stacked domain with only small differences can be found at 4.5s between crossline 7590 to 7790.

In addition, since our strategy of generating training pairs uses the last unblended shot gathers acquired from the edges of the survey, an analysis of the difference in the quality of the deblended output versus the distance of tested shots from the training data is therefore given as well. Figure 19 shows the NRMS of the differences between the input and the outputs of the proposed approach and the conventional algorithm, respectively, from 3.0 to 5.0 s TWT versus shot number of the above used 4000 shot gathers. The NRMS of the difference between the input and the output of the proposed approach can be smaller or larger than that of the conventional algorithm among different shot numbers, but in general, the performance of the proposed approach follows the overall trend of the conventional algorithm. For the entire tested sail line, the proposed approach was not influenced much by the distances between the test data and the training data.

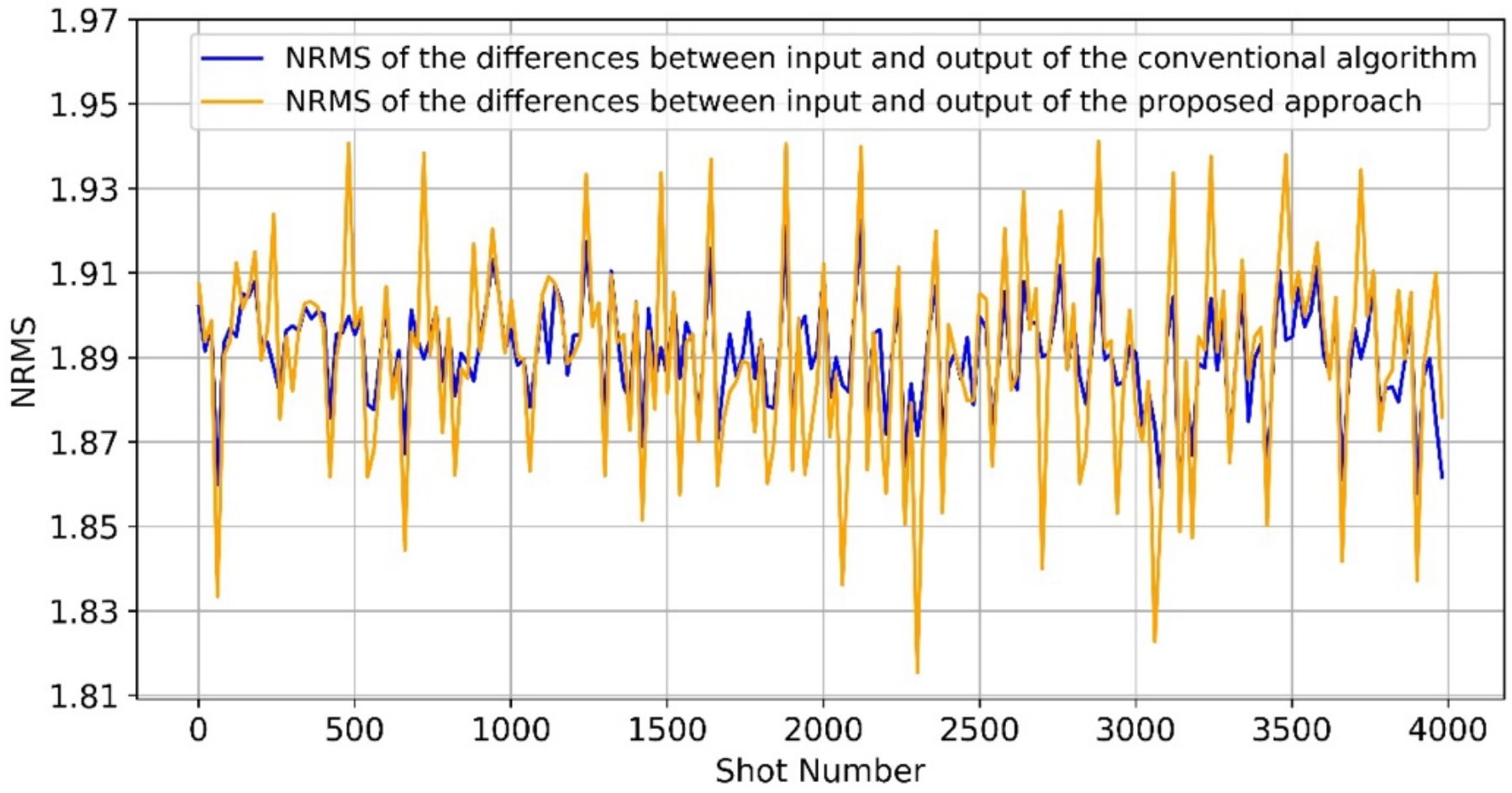

Figure 19: NRMS of the differences between input and outputs of the proposed approach and the conventional algorithm, respectively, from 3.0 to 5.0 s TWT versus shot number of a sail line.

## DISCUSSION

### Generalization of the proposed strategy to generate training data

In this study, we have proposed to use the last shot gathers from all streamers in different sail lines as the ground truth. The later samples of these gathers are always unblended with long enough recording length independent of the configuration of the actual blended acquisition. Existing methods either generate the ground truth from the application of a conventional deblending algorithm to a small part of the blended-by-acquisition data or alternatively from the use of synthetic data. In comparison, our strategy has significant advantages regarding both in generalization as well as in saving time and labor costs. The last shot gathers are easily obtained from the survey and can be directly accessed after completion of the blended acquisition. In addition, this strategy neither relies on available conventionally acquired seismic data nor the feasibility of using a DNN model trained by another survey. By manually blending shot gathers from the last shootings, we can therefore obtain high-quality training pairs with good control of the ground truth and fully adapted to the given survey area.

**Deblending in the shot domain**

The use of a small random jitter between the firing of shots is characteristic for a blended acquisition. This makes the application of denoising-based deblending methods feasible after resorting data from shot domain to an alternative domain like common-receiver or common-offset where the blending noise appears incoherent. The alternative approach of deblending data directly in the shot domain is more challenging due to the coherent character of the blending noise closely resembling the primary-source events.

The results obtained using the proposed DL approach are encouraging and demonstrate the feasibility of deblending directly in the acquisition domain, at least as a fast-track result. However, training based on data fired at the end of sail lines excludes real-time processing or quality control onboard seismic vessels during acquisition. As an alternative to the use of the last shot gathers as ground truth, using unblended shot gathers acquired from a previously conducted conventional survey or conducting a short period of unblended acquisition in the early stage of a blended survey could also be considered.

**Effectiveness of the data conditioning steps**

Among the proposed data conditioning steps, the most significant improvement was obtained when a multi-channel input was employed by adding adjacent blended shot gathers to the to-be-deblended input. A notable reduction in leakage of primary-source events was observed when compared to a single-channel input. This step is also the most computationally demanding among the proposed ones. The other two conditioning steps, e.g. predicting both primary-source events and (scaled-down) blending noise, and to align the to-be-deblended shot gathers by blending noise, gave only minor improvements when employed separately. However, the combined use of these steps with the multi-channel input led to visible improvement in the performance of our network. This is clearly demonstrated in the field data section.

**Difference between the training and test data sets**

Adjacent field blended shot gathers were employed as additional channels in the test data used for the initially trained network. However, no consecutive unblended shot gathers existed that could be used as multi-channel input data for the training process. We therefore employed adjacent streamers within the last shot gathers in a sail line as channels of a training sample. The distance between two adjacent streamers was 62.5 m. This is much larger than the distance between two adjacent blended shot gathers from the same survey being 37.5 m. Thus, dipping events will appear differently in a training sample compared to a test sample. One possible solution to this problem could be to acquire a set of consecutive unblended shot gathers in the beginning of a blended survey as already mentioned. Alternatively, a better loss function or the application of a NMO correction before the input samples were generated, may also have the potential to improve the network performance.

**Computational efficiency of the proposed approach**

It is always very challenging to extract and preserve the relatively very weak primary-source energy while effectively removing strong blending noise. In order to achieve such good deblending results as shown in Figure 13, a workflow of multiple processing steps were employed. Thus, the application of such a conventional deblending algorithm is computationally demanding and also time-consuming due to a large amount of parameter tuning. The level of deblending can vary considerably from one project to another, being dependent on both the number and position of sources, frequency content of interest, water depth among others. As a consequence, different deblending approaches and workflows may vary from project to project.

In contrast, the proposed DL approach has significant advantages in saving labor costs and computing time. In this study, the training process of the proposed approach took around 5.5 hours. Once finished the training, it took less than 0.3 s to deblend a single shot gather at the application stage. The runtime of the conventional algorithm used in this study is more than an order of magnitude higher than the complete route of the proposed approach including training.

## CONCLUSION

In this study, we have investigated a DL approach to seismic deblending in the shot domain and demonstrated its feasibility. A new strategy was proposed where the training data were manually blended by the unblended shot gathers acquired at the end of each sail line. After the DNN was properly trained, it efficiently removed blending noise from blended-by-acquisition data in the shot domain. Successful implementation of this method also demonstrates how unblended shot gathers can be used to train a DNN in the acquisition domain in a generic sense. As an alternative, unblended shot gathers could also be extracted from previous surveys, or deliberately acquired as part of the blended seismic acquisition.

The DNN performed especially well in the shallower part of the blended section but showed more leakage of primary-source events for larger traveltimes. However, by introducing a set of data conditioning steps, this leakage was reduced considerably. Direct comparison with an advanced conventional algorithm further demonstrated that the proposed approach gave slightly weaker primary-events and loss of high frequencies for the deep part of the same blended section. However, it is important to remember that such a conventional workflow has been developed and refined over years and in the case shown here this workflow consisted of multiple sub-processing steps and was very computationally demanding. Despite the present shortcomings, this study has demonstrated the data-driven nature and learning potential of a modern DNN when employed to deblending of seismic data and also specified our direction for future research.